\documentclass[twocolumn,showpacs,preprintnumbers,amsmath,amssymb]{revtex4}
\usepackage{graphics}
\usepackage{dcolumn}
\usepackage{bm}

\begin{document}
%


\title{Representation of SO(3) Group by a Maximally Entangled  State}
\author{W. LiMing} \email{wliming@scnu.edu.cn}
\author{Z. L. Tang}
\affiliation{Dept. of Physics, South China Normal University,
Guangzhou 510631, China}
\author{C. J. Liao}
\affiliation{School for Information and Optoelectronic Science and
Engineering, South China Normal University, Guangzhou 510631,
China}

\date{\today}

\begin{abstract}
A representation of the SO(3) group is mapped into a maximally
entangled two qubit state according to literatures. To show the
evolution of the entangled state, a model is set up on an
maximally entangled electron pair, two electrons of which pass
independently through a rotating magnetic field. It is found that
the evolution path of the entangled state in the SO(3) sphere
breaks an odd or even number of times, corresponding to the double
connectedness of the SO(3) group. An odd number of breaks leads to
an additional $\pi$ phase to the entangled state, but an even
number of breaks does not. A scheme to trace the evolution of the
entangled state is proposed by means of entangled photon pairs and
Kerr medium, allowing observation of the additional $\pi$ phase.
\end{abstract}
\maketitle

It is well known that when the spin of a spin-$\frac 1 2$ particle
rotates for a whole cycle on the Bloch sphere the wave-function of
the particle changes a phase of $\pi$. This $\pi$ phase has been
observed in several experiments\cite{Werner,Rauch}. This property
is commonly attributed to the topological property, i.e., the
double connectedness of the SO(3) group. The path on the manifold
of the SO(3) group is categorized into two classes under a
continuous deformation, one of which leads to a change of $\pi$ in
phase to the wave function, the other does not. However, it was
argued by Milman and Mosseri\cite{Milman} that this $\pi$ phase
may be shared by the multi-connectedness of both SO(3) and SO(2)
groups. They also argued that, in general, the $\pi$ phase is
partly geometric and partly dynamic. Only in the extreme case that
the spin precesses on the $xy$ plane in the Bloch sphere the $\pi$
phase is fully geometric. Especially, the $\pi$ phase still exists
when the spin initially points to the same direction of the
magnetic field, where there is no rotation at all. In such case
the $\pi$ phase is fully dynamic. Therefore, this $\pi$ phase may
not be directly related to the SO(3) group.

Milman and Mosseri found a one-to-one correspondence between the
representation of the SO(3) group and the evolution of a maximally
entangled state of a two-qubit system(MES)\cite{Milman}. They
adopted a discontinuously changing magnetic field, which suddenly
jumps from one direction to another. This is hardly possible to be
accomplished in reality. In the present paper a rotating magnetic
field is used to drive the evolution of a MES. A clearer formalism
is presented for the trajectory in SO(3).

A MES finds great application to quantum communication and quantum
computation techniques, and also to the study of fundamental
problems, e.g., non-locality, of quantum mechanics\cite{Buttler,
Zeilinger, Pan}. Much attention has been paid to MES's in recent
years. It is interesting that MES can be applied to the
representation of the SO(3) group.

\section{Mapping between a MES and SO(3)}

A two-qubit maximally entangled state (MES) of a two-state system
can be written as\cite{Milman},
\begin{equation}
|(\alpha,\beta)\rangle = \frac {1} {\sqrt 2} (\alpha|00\rangle
+\beta|01\rangle - \beta^*|10\rangle + \alpha^* |11\rangle)
\label{ab}
\end{equation}
where the coefficients $\alpha$ and $\beta$ are normalized to
unity
\begin{equation}
\alpha\alpha^*+\beta\beta^* = 1.
\end{equation}
It is seen that a MES is defined by a pair of complex numbers
$(\alpha, \beta)$. To visualize a MES, $\alpha$ and $\beta$ can be
parameterized to
\begin{eqnarray}
\label{alpha}
\alpha &=& \cos\frac a 2 - ik_z \sin \frac a 2,  \\
\label{beta} \beta  &=& -(k_y + ik_x) \sin \frac a 2,
\end{eqnarray}
where $(k_x, k_y, k_z)={\bf k}$ is a unit vector, and $a$ is an
angle between $0$ and $\pi$. Hence a MES can also be written as
$|\Psi({\bf k},a)\rangle$ in the parameter space. It is easy to
check that $|\Psi({\bf k}, \pi+a)\rangle = -|\Psi(-{\bf k},
\pi-a)\rangle$. That is, $({\bf k}, \pi+a)$ and $({\bf k}, \pi-a)$
correspond to the same state except for a global phase factor.
This is just the case of the double-valued representation of the
SO(3) group, which is written as
\begin{equation}
\label{Dka}
D^{1/2}({\bf k},a) = \binom { \alpha \quad\quad \beta}
{-\beta^* \quad \alpha^*},
\end{equation}
corresponding to a rotation $R({\bf k}, a)$ in real space to a
two-state particle. Although $R({\bf k}, \pi+a)$ and $R(-{\bf k},
\pi-a)$ are the same rotation, one has $D^{1/2}({\bf k},
\pi+a)=-D^{1/2}(-{\bf k}, \pi-a)$.
Therefore, there is a one-to-one correspondence between the
two-qubit MES and the double-valued representation of SO(3). In
fact, Any MES can be constructed by a rotation from an initial
MES, e.g.,
\begin{eqnarray}
\label{p10} D_1 |(1,0)\rangle &=& |(\alpha,\beta)\rangle,\\
 D_1 &\equiv & D^{1/2}_1({\bf k},a) \nonumber
\end{eqnarray}
where $D_1$ operates on the first particle. If a rotation operates
on the second particle, one has
\begin{eqnarray}
 D_2 |(1,0)\rangle &=& |(\alpha,-\beta^*)\rangle,\\
 D_2 & \equiv & D^{1/2}_2({\bf k},a) \nonumber
\end{eqnarray}

One could define a SO(3) sphere with diameter $\pi$ filled by
vectors $a{\bf k}=(ak_x, ak_y, ak_z)$. Due to (\ref{p10}) a MES
corresponds to a point in the SO(3) sphere, and an evolution of
MES corresponds to a trajectory connecting two points. The initial
state $|(1,0)\rangle$ locates at the center of the SO(3) sphere.

\section{A model Hamiltonian}

 Consider an electron in a rotating
magnetic field ${\bf B}(t) = B(\sin\theta \cos\omega t, \sin\theta
\sin\omega t, \cos\theta), $ where $\theta$ is the angle between
the field and the z-axis, and $\omega$ is the rotating frequency
of the field. The Hamiltonian of the electron is given by
\begin{eqnarray}
H(t) &=& {\bf \hat{\sigma} \cdot B}(t) \nonumber \\ &=& B \binom
{\cos\theta \,\,\,\,\,\,\,\, \sin\theta e^{-i\omega t}}
{\sin\theta e^{i\omega t}\,\,\,\,\,\,\,\, -\cos\theta \,\,\,\,}.
\end{eqnarray}
The two exact solutions of the time-dependent Schr\"{o}dinger
equation are given by
\begin{equation}
 |\psi_\pm (t)\rangle = \binom {a_\pm e^{-i\omega t/2}}
  {b_\pm e^{i\omega t/2}}e^{\mp i \omega_0 t},
\end{equation}
corresponding to energy eigenvalues $\hbar\omega_0$ and
$-\hbar\omega_0$, respectively, where
\begin{eqnarray}
b_\pm &=&a_\pm\frac {\hbar (\omega \pm 2\omega_0)-2B \cos\theta} {2 B \sin\theta},\\
\omega_0 &=&\frac 1 {2 \hbar} \sqrt{(\hbar\omega)^2 -4B\hbar\omega
\cos\theta+ 4B^2},
\end{eqnarray}
where the values of $a_\pm$ can be determined by normalization of
solutions.

 Now consider an initial state $|(1,0)\rangle =(|00\rangle +
|11\rangle)/\sqrt 2$ of two electrons, where $|0\rangle =
|\psi_+(0)\rangle, |1\rangle = \psi_-(0)\rangle$. Suppose the
first electron travels through a rotating magnetic field, then the
system of two electrons evolutes in the form
\begin{eqnarray}
|(1,0)\rangle &\rightarrow & [|\psi_+(t)\psi_+(0)\rangle
\nonumber \\
&+&|\psi_-(t) \psi_-(0)\rangle]/\sqrt(2) \\
\label{evolution1}
 &=& |(\alpha, \beta)\rangle=D_1(\omega t,\omega_0) |(1,0)\rangle
\end{eqnarray}
where new arguments have been assigned to the group element for
convenience, and
\begin{eqnarray}
\label{a1}
 \alpha &=&[ \cos\frac {\omega t} 2 + i\sin\frac {\omega
t} 2 \frac {\hbar \omega -2B \cos\theta} {2 \hbar \omega_0} ]
e^{-i\omega_0 t}, \\
\label{b1}
\beta &=&i \sin\frac {\omega t} 2 \frac {a_+} {a_-}
\frac {\hbar (\omega-2\omega_0)-2B \cos\theta} {2\hbar \omega_0}
e^{i\omega_0 t}.
\end{eqnarray}
It is seen that a rotating magnetic field leads to an evolution of
a MES through a continuous trajectory in the SO(3) sphere.
Therefore, a rotating magnetic field is equivalent to a three
dimensional rotation in real space to the MES.

It is not surprising that when $\omega t=2\pi, \omega_0=n\omega,
n={\rm integers}$, the initial state acquires an additional phase
of $\pi$, i.e.,
\begin{equation}
D_1(2\pi,\omega) |(1,0)\rangle = -|(1,0)\rangle
\end{equation}
The amazing property is that the above operation can be allocated
to two particles of the initial state, i.e.,
\begin{equation}
D_1(\pi,n\omega) D_2(\pi,n\omega)|(1,0)\rangle = -|(1,0)\rangle
\end{equation}
In general, one does not have such a property for other initial
state. If $\omega_0=(n+1/2)\omega, n={\rm integers}$ one will have
\begin{equation}
D_1(\pi,\omega_0) D_2(\pi,\omega_0)|(1,0)\rangle = |(1,0)\rangle,
\end{equation}
acquiring no additional phase. Hence, one has a choice for the
additional phase through selecting the value of $\omega_0$.

Now we can trace the following evolution
\begin{eqnarray}
\label{evolution}
|(1,0)\rangle &\rightarrow &
[|\psi_+(t)\psi_+(t)\rangle+|\psi_-(t) \psi_-(t)\rangle]/\sqrt(2) \\
 &=& D_1(\omega t,\omega_0)
 D_2(\omega t,\omega_0) |(1,0)\rangle
\end{eqnarray}
Under the choice $\omega_0=n\omega, \,{\rm or} \, (n+1/2)\omega,
n={\rm integers}$ this evolution makes a closed trajectory in the
SO(3) sphere. An example is shown in Fig.1, where parameters
$\theta$ and $B$ are set to meet $\omega_0=\omega$. The final time
is $t=\pi/\omega$, that is, Both magnetic fields of the two
electrons rotate half a cycle. It is seen that this trajectory
breaks three times on the surface of the sphere. It is known that
two ends of a diameter of the sphere correspond to the same
rotation but the group element, (\ref{Dka}), changes its sign. In
this case, through a whole trajectory, the MES acquires an
additional phase of $\pi$.

\begin{figure}
\includegraphics{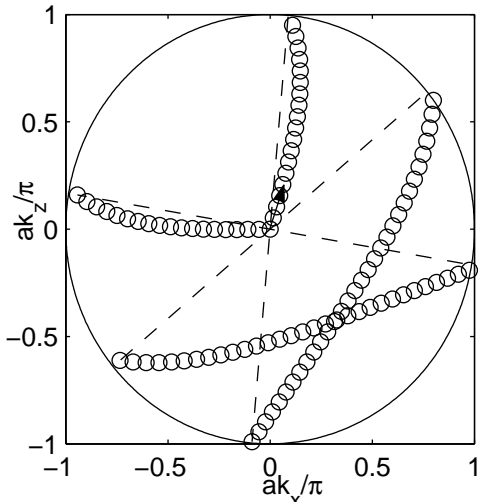}
\caption{Closed trajectory in the SO(3) sphere with $\theta=\frac
\pi 5, B = 1.3603$. The arrow stands for the beginning state and
direction of evolution.}
\end{figure}

With proper parameters, one can have closed trajectories with even
numbers of breaks, corresponding to a change of $2\pi$ in phase.
An example is shown in Fig.2. Hence, one has two classes of
trajectories, one of which has an odd number of breaks on the
surface of the SO(3) sphere and the other has an even number of
breaks, corresponding to the two classes of the double
connectedness of the SO(3) group. This is the case that Milman and
Mosseri considered\cite{Milman}, whereas their trajectories are
hardly possible to be realized, since their magnetic field has to
jump through a few discrete points in the parameter space.

\begin{figure}
\includegraphics{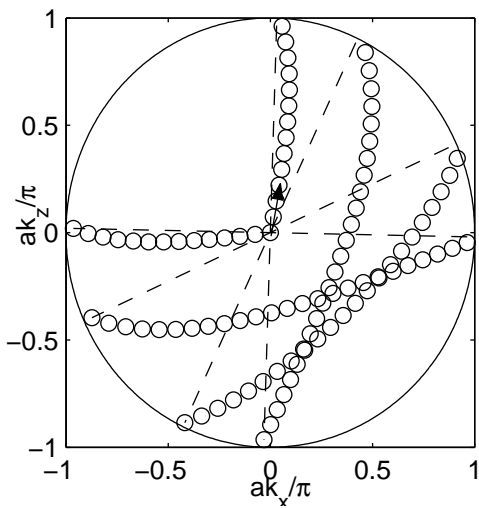}
\caption{Closed trajectory in the SO(3) sphere with $\theta=\frac
\pi 5, B = 1.8754$. The arrow stands for the beginning state and
direction of evolution.}
\end{figure}

It can easily checked that the closed trajectories $A-B-F-D-A$ and
$A-B-F-\bar E-\bar A$ and other ones that Milman and Mosseri
considered \cite{Milman} belong to the two simplest classes of
trajectories which have $0$ or $1$ breaks, respectively, in the
SO(3) sphere. Therefore, the present work extends their model to
include a great number of closed trajectories with even or odd
numbers of breaks.

\section{Realization of the $\pi$ phase by entangled photon pairs}

A entangled photon pair emerging from a double refraction
crystal\cite{Kwiat} can be in one of four Bell states
$|\Phi^+\rangle = (|H_a H_b\rangle + V_a V_b\rangle)/\sqrt 2$,
where $|H\rangle$ and $|V\rangle$ denote a horizontally polarized
photon state and a vertically polarized one , respectively. This
two entangled photons separate with each other after emission, and
then pass through two negative Kerr medium P1 and P3, and two
positive Kerr medium P2 and P4, as seen in Fig.3. The Kerr medium
are modulated by electric fields, so that their optical axes are
in directions as shown in the lower part in Fig.3. P1 and P3 point
to the same direction, say z-axis, and the directions of P2 and P4
can be adjusted by changing the directions of their electric
fields.

\begin{figure}
\includegraphics{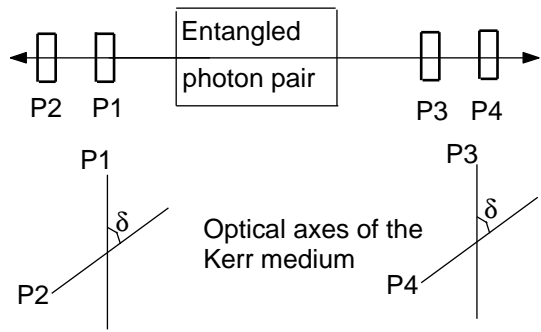}
\caption{Scheme to produce the $\pi$ phase. P1, P2, P3 and P4 are
Kerr medium. The optical axes are given by the directions of the
electric fields on the Kerr medium.}
\end{figure}

P1 and P3 change the relative phase between $|H\rangle$ and
$|V\rangle$, working as the following matrix
\begin{eqnarray}
\label{U1}
U_1&=&\binom {e^{-i\phi_1/2} \quad \quad 0} {0 \quad \quad e^{i\phi_1/2}},\\
\phi_1&=&\frac {2 \pi} \lambda (n_{e1} - n_{o1})d_1 = \frac {2
\pi} \lambda k_1 d_1 E_1^2,
\end{eqnarray}
where $E_1$ is the electric field applied to Kerr medium P1 and
P3. Since the optical exes of P2 and P4 take an angle of $\delta$
with the z-axis, they work as the following matrix,
\begin{eqnarray}
U_2&=&\binom {A\quad\quad  B} {-B^* \quad A^*}, \\
\label{A}
A&=& \cos\frac {\phi_2} 2 + i\sin\frac {\phi_2} 2 \cos {2\delta} ,  \\
\label{B}
B&=&i\sin\frac {\phi_2} 2 \sin {2\delta} ,  \\
\phi_2 &=& \frac {2 \pi} \lambda (n_{o2} - n_{e2})d_2 = \frac {2
\pi} \lambda k_2 d_2 E_2^2,
\end{eqnarray}
where $E_2$ is the electric field applied to P3 and P4.

It is seen that the combination, $U_2U_1$, is equivalent to a
rotating magnetic field. By comparing
eqs.(\ref{evolution1},\ref{a1},\ref{b1}) and
eqs.(\ref{U1},\ref{A},\ref{B}) one finds correspondence $\phi_1
\sim \omega_0 t, \phi_2 \sim \omega t, \cos2\delta \sim \frac
{\hbar \omega -2B \cos\theta} {2 \hbar \omega_0}$. Hence, the
evolution in eq.(\ref{evolution}), as shown in Fig.1 and Fig.2,
can be exactly traced by varying the electric fields $E_1$ and
$E_2$ on the Kerr medium. With proper values of electric fields
such that $\phi_1=n\phi_2$ one may obtain an additional $\pi$
phase, or zero additional phase if $\phi_1=(n+\frac 1 2)\phi_2$.

The $\pi$ phase can be easily observed by various interference
experiments. For example, according to the scheme described by
Milman and Mosseri considered \cite{Milman}, one arm of the
entangled photon pair can be transformed into a Mach-Zender
interferometer. The two wave plates in that scheme are replaced by
combinations of Kerr medium P1 and P2, and that of P3 and P4.

In summary, the present paper sets up a representation for the
SO(3) group by maximally entangled two-qubit states. The evolution
of the entangled states showed the double connectedness of the
SO(3) group. In the SO(3) sphere the evolution path breaks an odd
or even number of times. An odd number of breaks causes an
additional $\pi$ phase to the entangled state, but an even number
of breaks does not. The additional $\pi$ phase can be observed by
interference experiments of entangled photon pairs.

\acknowledgments{The author acknowledges Shidong Liang for helpful
discussions and financial support of the Science and technology
project of Guangzhou 2001-2-095-01.}

\end{document}